\documentclass[12pt]{iopart}
\usepackage{graphicx,latexsym,stmaryrd,iopams}

\usepackage{color}

\begin{document}

\title{Exact results for the $1$D interacting Fermi gas with arbitrary polarization}

\author{M.T. Batchelor$^{1,2}$, M. Bortz$^1$, X.W. Guan$^{1,2}$ and N. Oelkers$^{1,2}$}
\address{$^1$
{\small Department of Theoretical Physics, Research School of Physical Sciences and Engineering}}
\address{$^2$
{\small Department of Mathematics, Mathematical Sciences Institute\\}
{\small The Australian National University, Canberra ACT 0200, Australia}}


\begin{abstract}
We investigate the $1$D interacting two-component Fermi gas with arbitrary polarization. 
Exact results for the ground state energy, quasimomentum distribution functions, spin velocity and
charge velocity reveal subtle polarization dependent quantum effects.  
\end{abstract}


\maketitle

Recent achievements in trapping quantum gases of ultra-cold atoms have
opened up many exciting possibilities for the experimental
investigation of quantum effects in low-dimensional many-body systems.
Among these, progress in realizing interacting Fermi gases 
promises new opportunities for studying the pronounced crossover from a
Bardeen-Cooper-Schrieffer (BCS) superfluid to a Bose-Einstein
condensate (BEC) \cite{BEC-F}.
A crossover from weakly attractive Cooper pairs  to strongly repulsive 
molecular dimers can also take place in a confined quasi $1$D Fermi gas of atoms  
by using magnetic field induced Feshbach resonances \cite{Fermi-1D1,Fermi-1D2}. 
In this scenario, a number of key $1$D integrable models with a
variable interaction parameter have been used to describe the crossover 
\cite{Fermi-1D2,BEC-BCS2,BEC-BCS3,BEC-BCS1,BEC-BCS4,BEC-BCS5,BEC-BCS6}.
These include the integrable interacting Fermi gas \cite{Gaudin-Yang}.

The $1$D Fermi gas can be experimentally realized \cite{Fermi-1D1} by tightly confining 
the atomic cloud in the radial directions and weakly confining it along the axial direction
\cite{Fermi-1D2,BEC-BCS2,BEC-BCS3,BEC-BCS1,SC-S}. 
To date, the theoretical study of the $1$D continuum Fermi gas has been mainly 
for the zero polarization case in the thermodynamic limit (TL).
Our aim here is to present the results of a comprehensive study of the $1$D
exactly solved two-component Fermi gas with arbitrary polarization, throughout
the whole interaction range. 
This includes finite systems in the weak and strong coupling regions, and in the TL.

{\it The model.} 
The Hamiltonian \cite{Gaudin-Yang}   
\begin{equation}
{H}=-\frac{\hbar ^2}{2m}\sum_{i = 1}^{N}\frac{\partial
^2}{\partial x_i^2}+\,g_{\rm 1D} \sum_{1\leq i<j\leq N} \delta
(x_i-x_j)
\label{Ham-1}
\end{equation}
describes a $\delta$-function interacting gas of $N$ fermions each of mass $m$
constrained by periodic boundary conditions to a line of length $L$.
The coupling constant $g_{\rm 1D}$ can be written in terms of 
the scattering strength $c={2}/{a_{\rm 1D}}$ as $g_{\rm 1D} ={\hbar ^2 c}/{m}$.
An effective $1$D scattering length $a_{\rm 1D}$ can be expressed through the 
$3$D scattering length for fermions confined in a $1$D geometry. 
We use the dimensionless coupling constant $\gamma={mg_{\rm 1D}}/{(\hbar^2n})$ 
for physical analysis. 
Here $n={N}/{L}$ is the linear density. 
The coupling $g_{\rm 1D}$ can vary from $-\infty$ to $\infty$ by 
detuning the $3$D scattering length.
The interaction is attractive for $g_{\rm 1D}<0$ and repulsive for $g_{\rm 1D}>0$.

The energy is given by $E=\frac{\hbar ^2}{2m}\sum_{j=1}^Nk_j^2$, where 
the quasimomenta $k_j$ satisfy the Bethe equations (BE) \cite{Gaudin-Yang}
%
%
%
\begin{eqnarray}
& &\exp(\mathrm{i}k_jL)=\prod^M_{\ell = 1} 
\frac{k_j-\Lambda_\ell+\mathrm{i}\, c/2}{k_j-\Lambda_\ell-\mathrm{i}\, c/2},\nonumber\\
& &\prod^N_{\ell = 1}\frac{\Lambda_{\alpha}-k_{\ell}+\mathrm{i}\, c/2}{\Lambda_{\alpha}-k_{\ell}-\mathrm{i}\, c/2}
 = - {\prod^M_{ \beta = 1} }
\frac{\Lambda_{\alpha}-\Lambda_{\beta} +\mathrm{i}\, c}{\Lambda_{\alpha}-\Lambda_{\beta} -\mathrm{i}\, c} .
\label{BE}
\end{eqnarray}
Here $j = 1,\ldots, N$ and $\alpha = 1,\ldots, M$, with $M$ the number of spin-down fermions.
For convenience we take $N$ even.
Of particular interest are the bound states for attractive interaction. 
In this regime, the model (\ref{Ham-1}) has been extensively studied with $N=2M$ 
in the context of the BCS-BEC crossover \cite{Fermi-1D2,BEC-BCS2,BEC-BCS3,BEC-BCS1,BEC-BCS4}.  
For weak attraction, the system describes weakly bound Cooper pairs. 
In the strongly attractive regime the bound states behave like bosonic 
tightly bound molecular dimers. 
On the other hand, for repulsive interaction, there are no bound states for the ground state. 
The spin-up and spin-down fermions behave like indistinguishable free fermions
as $c\to \infty$. 
The ground state energy per particle in the TL,  i.e., $N, M, L \to \infty$, $N/L, M/L$ constant, 
can be derived from the Gaudin integral equations \cite{Gaudin-Yang}. 
Here we first analyze the BE directly, i.e., for a finite system.

{\it The ground state.}
We observe that for weak attractive interaction a pair of fermions with spin-up and 
spin-down states forms a bound state. 
On the other hand, for weak repulsive interaction the fermion momenta 
separate  in quasimomentum space. 
For $L|c| \ll 1$, it is convenient to distinguish between unpaired 
$k_j^{(\rm u)}$ ($j=1,\ldots,N-2M$) and paired $k_j^{(\rm p)}$ ($j=1,\ldots,2M$) momenta. 
On expanding the BE in powers including $\mathcal{O}(c)$ \cite{BGM}, 
one obtains $k_j^{(\rm u)}=\pi(M-N-1+2j)/L+\delta_j^{(\rm u)},\,j=1,\ldots,N/2-M$, $k_j^{(\rm u)}=\pi(3M-N-1+2j)/L+\delta_j^{(\rm u)},\,j=N/2-M+1,\ldots,N-2M$,  and 
$k_j^{(\rm p)}=\pi \left(-1-M+2j_+\right)/L+ \delta^{(\rm  p)}_{j_+}\pm \sqrt{c/L}$, $(j=1,\ldots,M)$,
where $j_+=j$, if $j$ odd and $j_+=j-1$ if $j$ even. Therefrom it follows that the binding energy in the weakly attractive case is $c/L$ for finite systems.
The deviations $\delta$ from $k_j^{(\rm u,p)}$ are linear in $c$, with
\begin{eqnarray}
\delta_j^{(\rm u)}&=&\frac cL {\sum_\ell}' \frac{1}{k_{j,0}^{(\rm u)}
-k_{\ell,0}^{(\rm p)}},\label{del1}\\
\delta_j^{(\rm p)}&=&\frac cL \left[{\sum_\ell}' \frac{1}{k_{j,0}^{(\rm p)}
-k_{\ell,0}^{(\rm p)}}+\frac{1}{2}\sum_{\ell}\frac{1}{k_{j,0}^{(\rm p)}
-k_{\ell,0}^{(\rm u)}}\right]\label{del2},
\end{eqnarray}
where the primed sum counts each $k_{\ell,0}^{(\rm p)}$ only once and excludes $k_{j,0}^{(\rm p)}$.
Here $k_{j,0}^{(\rm u,p)}$ are the quasimomenta of non-interacting particles 
as given above.  
Note that the unpaired momenta do not interact with each other, which is consistent with the Pauli principle.  
On the other hand, for strong attractive interaction, i.e., $Lc\ll -1$,
tightly bound states with quasimomentum 
$k_i^{(\rm p)} \approx \Lambda_i \pm \frac{1}{2} {\mathrm i} c$  
separate from unbound states with quasimomentum
$k_j^{(\rm u)}=\frac{n_j\pi}{L}\left(1+\frac{4M}{Lc}\right)^{-1}$ with integers 
$n_j=\pm 1, \pm 3,\ldots, \pm(N-2M-1)$. 
In the above, $\Lambda_i=\frac{n_i\pi}{L}\left(1+\frac{M}{Lc}+\frac{2(N-2M)}{Lc}\right)^{-1}$
for $n_i=\pm 1,\ldots, \pm \frac12{(M-1)}$. 
In this scenario the bound states behave like
hard-core bosons due to Fermi pressure.

We now define the fraction $a={2M}/{N}$ which characterizes the polarization of the 
system.\footnote{The polarization dependence of atom-hole condensation in optical lattices was discussed
in Ref.~\cite{Lee}.}
The ground state energy per particle in the weakly interacting regime follows from the 
above asymptotic roots of the BE as (in units of $\hbar^2n^2/(2m)$, with 
$a_0=(1-a)\left((1-\frac{1}{2}a)^2+\frac{1}{2}a\right)$)
\begin{eqnarray}
\frac{E}{N}&=& 
\frac{a^2}{2}\gamma+\gamma(1-a)a+\frac{\pi^2}{12}a^3+\frac{\pi^2}{3}a_0
+\mathcal{O}(\gamma^2).\label{E-W}
\end{eqnarray}
The first two terms represent the binding and the interaction energy, respectively, which are both linear in $\gamma$. 
This is different from what is obtained in the TL from the Gaudin
integral equations \cite{BEC-BCS2,BEC-BCS1,BEC-BCS4}, where the binding energy is $\propto \gamma^2$. 
The other terms in (\ref{E-W}) are the kinetic energies for paired and unpaired
fermions. 
In the special case $a=0$, the model reduces to fully polarized free fermions without $s$-wave scattering. 
Using the given asymptotic expressions for the Bethe roots in the strongly 
attractive regime, we obtain (in units of $\hbar^2n^2/(2m)$)
\begin{eqnarray}
\frac{E}{N}&=& 
-\frac{a}{4}\gamma^2+ \frac{\pi^2}{3} \left( \frac{a^3}{16} (1 - 2b_1) +
a_1(1-2b_2) \right) + \mathcal{O}(1/\gamma^2) , \label{E-S-A}
\end{eqnarray}
where $a_1=(1-a)^3$, 
$b_1 = a/(2\gamma) + 2(1-a)/\gamma$ and
$b_2 = 2a/\gamma$.
Whereas the first term represents the binding energy, the remaining interaction energies stem from
the interaction between pair-pair and pair-unpaired fermions.  
The ground state energy is lowest for $a=1$, compared to the others for $0\le a<1$. 
This is in general agreement with the Lieb-Mattis theorem \cite{LM-thorem}.

\begin{figure}[t]
\begin{center}
\includegraphics[width=.80\linewidth]{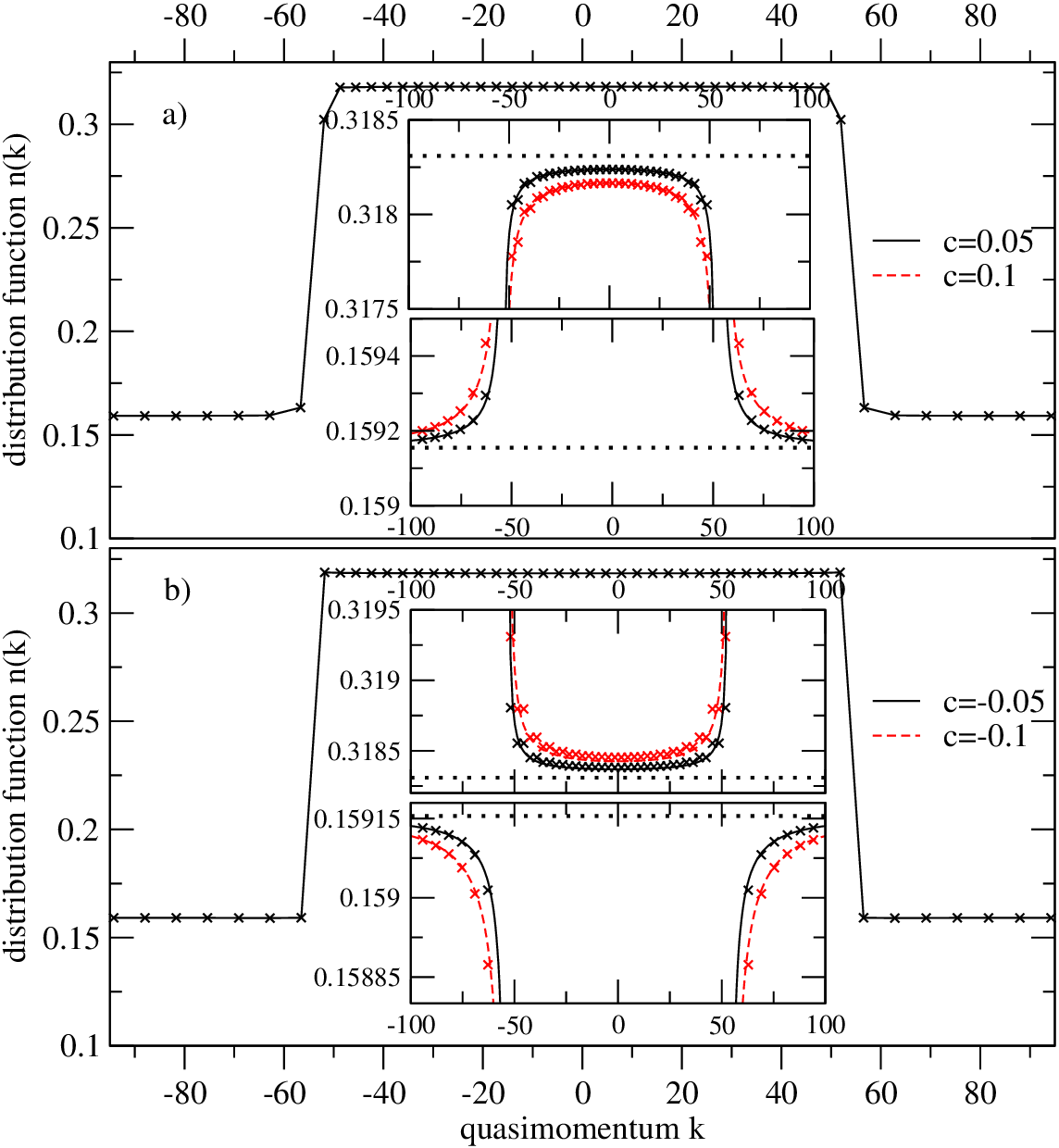}
\end{center}
\caption{
The quasimomentum distribution function (crosses) for a finite system ($N=50, M=16$) 
for the a) weakly repulsive and b) weakly attractive regime. 
The full lines connecting the crosses are guides to the eye. 
The insets are zooms of the distribution near the noninteracting cases 
(dotted lines), $1/2\pi$ for unpaired fermions and $1/\pi$ for paired fermions.
Here the full lines are the results obtained in the TL (Eqs.~\eref{n1}, \eref{n2}).
}
\label{Fig1}
\end{figure}

Now consider the crossover to the TL. 
First we observe that in the TL, the results \eref{E-W}, \eref{E-S-A} %
reduce to those derived
from the Gaudin integral equations \cite{BEC-BCS2,BEC-BCS3,BEC-BCS1,BEC-BCS4}.  
Furthermore, from \eref{del1}, \eref{del2}, we obtain the quasimomentum distribution functions 
$n_{\rm u,p}(k)$ per unit length in the TL, 
\begin{eqnarray}
\!\!\!n_u(k)&=& \frac{1}{2\pi} +\frac{c}{2\pi^2}\,\frac{A}{k^2-A^2},\; A<|k|<B,\label{n1}\\
\!\!\!n_p(k)&=& \frac1\pi -\frac{c}{2\pi^2}\left(\frac{A}{A^2-k^2}+\frac{B}{B^2-k^2}\right),\, |k|<A,\label{n2}
\end{eqnarray}
with $A=\pi M/L$ and $B=\pi (N+M)/L$. 
Especially, for $M=N/2$, $n_{\rm u}\equiv 0$.
In Fig. \ref{Fig1}, $n_{u,p}(k)$ as calculated from the finite BE are compared to Eqs. \eref{n1}, \eref{n2}.

\begin{figure}[t]
\begin{center}
\vspace{0.5cm}
\includegraphics[width=0.95\linewidth]{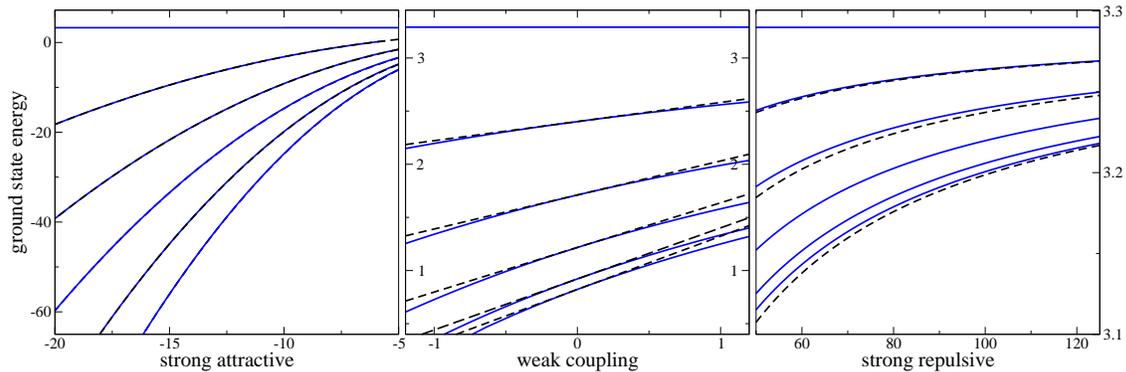}
\end{center} 
\caption{
Ground state energy in units of $\hbar^2 n^2/(2m)$ vs interaction strength $\gamma$ 
for different polarizations. 
Dashed lines compare the analytic approximations \eref{E-W}, \eref{E-S-A}, \eref{E-S-R} 
with the numerical solution of the integral equations (solid lines) in the 
different coupling regimes. 
The curves shown are for $a=0$, $0.2$, $0.4$, $0.6$, $0.8$, $1$ (top to bottom).
}
\label{Fig2}
\end{figure}

We now use the Gaudin integral equations to numerically compute the dependence of 
$E/N$ on $\gamma$ for different $a$. 
The result is shown in Fig. \ref{Fig2}. 
The numerical data agree with \eref{E-W} and \eref{E-S-A} in the weakly 
interacting and strongly attracting regimes. 
A straightforward analysis of the integral equations in the strongly repulsive regime leads to 
\begin{eqnarray}
\frac{E}{N}\approx \left\{\begin{array}{ll}
\frac{\hbar^2n^2}{2m}\frac{\pi^3}{3}\left(1 - \frac{4a}{\gamma}\right)
+\mathcal{O}(1/\gamma^2,a^2/\gamma),&\,a\ll1\\
 \frac{\hbar^2n^2}{2m}\frac{\pi^3}{3}\left(1-\frac{4 \ln 2}{\gamma}\right)
 +\mathcal{O}(1/\gamma^2),&\,a=1
\end{array}
\right.\label{E-S-R}
\end{eqnarray}
which provides a further check on the numerics.
As can be seen from \eref{E-S-R}, for strongly repulsive interaction the mixture behaves like a 
single component Fermi gas of indistinguishable particles,
rather than hard-core bosons \cite{BEC-BCS2,BEC-BCS3,BEC-BCS4}.

{\it Spin-charge separation.}
In the TL, the low-energy behaviour of interacting Bose and Fermi gases is
universal in the dispersion relation and correlation behaviour \cite{Haldane}. 
Spin-charge separation is one of the typical Luttinger liquid behaviours, with   
both the spin and charge degrees of freedom having different propagation velocities 
at Fermi surfaces.  
The charge-velocity can be calculated exactly from 
$v_c^2=\frac{L}{m n} \frac{\partial^2 E}{\partial L^2}$ \cite{London} 
for arbitrary $a$, which yields 
\begin{eqnarray}
\frac{v_c}{v_{\rm F}}=\left\{\begin{array}{ll}
\sqrt{\frac{a^3}{16} (1-4 b_1)+
a_1(1-4b_2)}, &\,\gamma \ll-1,\\
\sqrt{\frac{a(1-{a}/{2})\gamma}{\pi^2}+\frac{a^3}{4}+a_0},&\,|\gamma|\ll1,
\end{array}
\right.
\end{eqnarray}
where the Fermi velocity $v_{\rm F}={\hbar \pi n}/{m}$.

\begin{figure}[b]
\vspace{0.7cm}
\begin{center}
\includegraphics[width=0.80\linewidth]{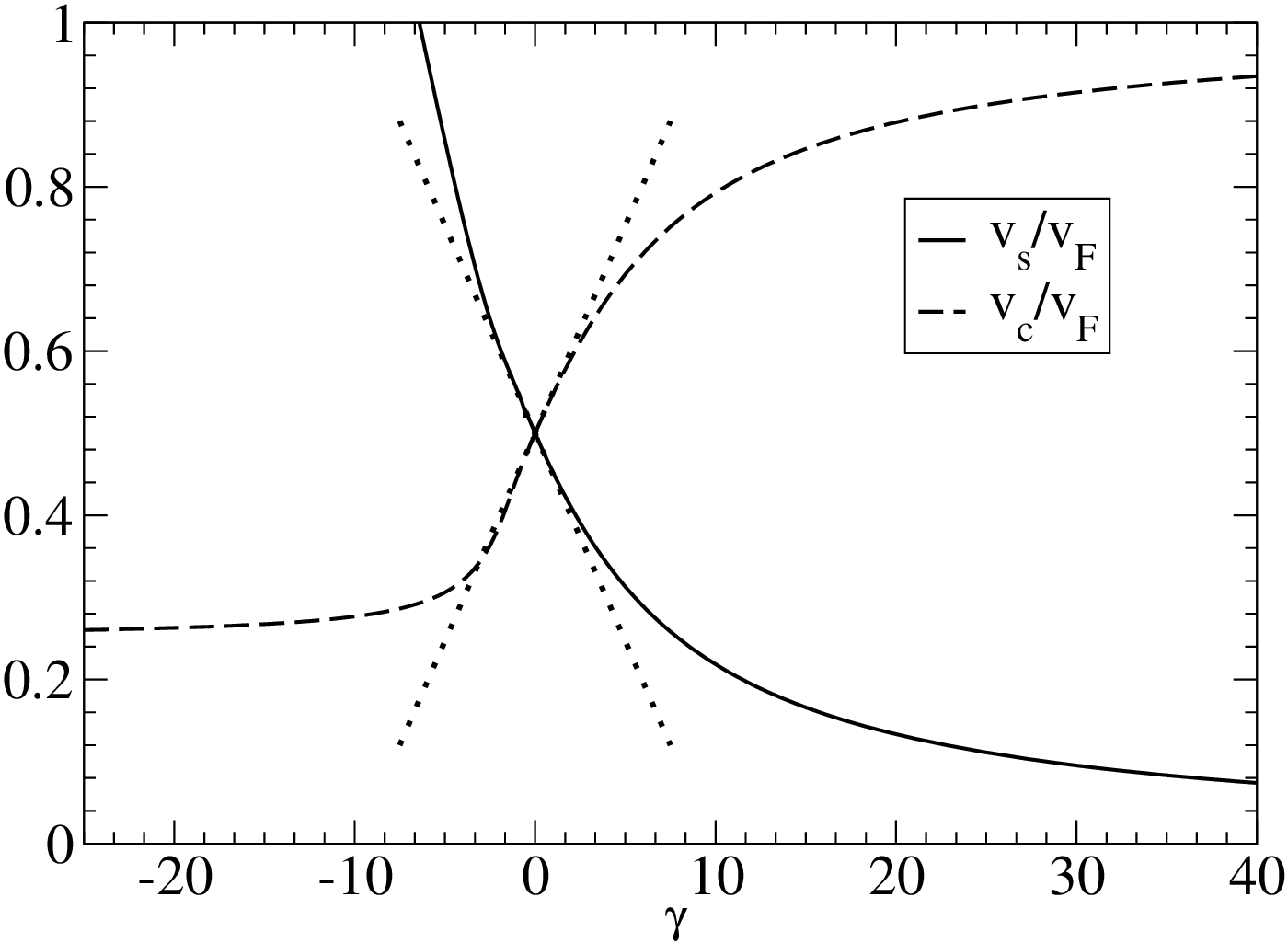}
\end{center}
\vspace{-1mm}
\caption{Spin and charge velocities vs interaction strength for $a=1$. 
The straight dotted lines show the lowest order in $\gamma$ from \eref{smg}, 
which is also accessible by field theory \cite{SC-S}.}
\label{Fig3}
\end{figure}

To illustrate spin-charge separation for non-zero interaction, we show the 
spin and charge velocities in Fig. \ref{Fig3} as a function of $\gamma$ for $a=1$. 
These curves are obtained numerically from the dressed-energy 
formalism \cite{Takahashi, Schlottmann}. 
For weak interaction \cite{BEC-BCS1}, 
\begin{eqnarray}
\frac{v_{c,s}}{v_{\rm F}}&=&\frac{1}{2}\left(1\pm\frac{\gamma}{\pi^2}\right), 
\quad |\gamma|\ll 1,\label{smg}
\end{eqnarray}
where the upper sign refers to $v_c$. 
Furthermore, $\frac{v_{c}}{v_{\rm F}}=1-\frac{4\ln 2}{\gamma}$, 
$\frac{v_s}{v_{\rm F}}=\frac{\pi^2}{3 \gamma}$ for $\gamma \to\infty$ and 
$\frac{v_{c}}{v_{\rm F}}= \frac{1}{4}(1-1/\gamma)$ 
for $\gamma \to -\infty$ \cite{BEC-BCS1}. 
In the attractive regime, the charge excitations are still gapless; 
however, a gap opens in the spin channel \cite{Krivnov}. 
This gap increases with increasing $|\gamma|$, leading to a divergent spin velocity in the 
strongly attractive limit, with $\frac{v_s}{v_{\rm F}}=-\frac{\gamma}{ 2\sqrt2 \pi}$ as 
$\gamma \to -\infty$ \cite{BEC-BCS1}. 
Some possible ways to experimentally observe the spin-charge separation in
harmonically trapped atoms have been suggested \cite{SC-S}.

In conclusion, we have examined in detail the exactly solved two-component
1D interacting Fermi gas for arbitrary polarization $a$ in the whole interaction range. 
Analytic results have followed for the quasimomentum density profile, the ground state energy and
the spin-charge velocities.
These quantities reveal significant insight into the nature of quantum effects in the Fermi gas
with arbitrary polarization.
 
\ack

This work has been supported by the Australian Research Council 
and the German Science Foundation (grant number BO/2538).
The authors also thank the organizers of the {\em Counting Complexity} meeting
in honour of Tony Guttmann's 60th birthday, where this work was presented.


\end{document}